\documentclass[aps,prl,preprint,groupedaddress]{revtex4-1}
\usepackage{graphicx,psfrag}
\usepackage{bm}
\usepackage{mathbbol,verbatim}
\usepackage{slashed}
\usepackage{graphics}
\usepackage{color,ulem, amsmath, amssymb}
\usepackage{tikz}
\usetikzlibrary{trees}
\usetikzlibrary{decorations.pathmorphing}
\usetikzlibrary{decorations.markings}
\usetikzlibrary{decorations, decorations.markings, decorations.pathmorphing, arrows, graphs, shapes.geometric, snakes}
\usetikzlibrary{arrows}
\usepackage[colorlinks=true,linkcolor=blue,citecolor=blue]{hyperref}%

\tikzset{
photon/.style={decorate, decoration={snake}, draw=red},
dark/.style={draw=gray, postaction={decorate},
        decoration={markings,mark=at position .55 with {\arrow[draw=gray]{>}}}},
antidark/.style={draw=gray, postaction={decorate},
        decoration={markings,mark=at position .55 with {\arrow[draw=gray]{<}}}},
electron/.style={draw=violet, postaction={decorate},
        decoration={markings,mark=at position .55 with {\arrow[draw=violet]{>}}}},
neutrino/.style={draw,color=violet,thick, postaction={decorate} },
quark/.style={draw=blue, postaction={decorate},
        decoration={markings,mark=at position .55 with {\arrow[draw=blue]{>}}}},
antiquark/.style={draw=blue, postaction={decorate},
        decoration={markings,mark=at position .55 with {\arrow[draw=blue]{<}}}},
gluon/.style={decorate, draw=or,
        decoration={coil,amplitude=2pt, segment length=3pt}},
ZZ/.style={decorate, decoration={snake,amplitude=1.5pt, segment length=5pt}, draw=greeen},
        left,
chargedscalar/.style={draw=black, postaction={decorate},
        decoration={markings,mark=at position .55 with {\arrow[scale=1.25,draw=black,>=latex]{>}}}},
  }

\definecolor{greeen}{rgb}{0.03,0.84,0.13}
\definecolor{test}{rgb}{0.03,0.74,0.33}
\definecolor{viol}{rgb}{0.44,0,0.94}
\definecolor{or}{rgb}{0.95,0.65,0}

\begin{document}
\begin{flushright}{UMD-PP-018-02}
\end{flushright}

\title{Natural Alignment of Quark Flavors and Radiatively Induced Quark Mixings}

\author{Abhish  Dev}
\author{and Rabindra N. Mohapatra}
\affiliation{Maryland Center for Fundamental Physics, Department of Physics, University of Maryland, College Park, MD 20742, USA}

\begin{abstract} The standard model does not provide an explanation of the observed alignment of quark flavors i.e. why are the up and down quarks approximately aligned in their weak interactions according to their masses? We suggest a resolution of this puzzle using a combination of left-right and Peccei-Quinn (PQ) symmetry. The quark mixings in this model vanish at the tree level and arise out of one loop radiative corrections which explain their smallness. The lepton mixings, on the other hand, appear at the tree level and are therefore larger. We show that all fermion masses and mixings can be fitted with a reasonable choice of parameters. The neutrino mass fit using seesaw mechanism requires the right-handed $W_R$ mass bigger than 18 TeV. Due to the presence of PQ symmetry, this model clearly provides a solution to the strong CP problem.
\end{abstract}

\maketitle

\section{1.  Introduction} The standard model (SM), in spite of its spectacular successes, leaves many questions unresolved such as the gauge hierarchy problem, the strong CP problem, neutrino masses, and dark matter. One rarely discussed puzzle is: why are the up and down quark flavors approximately aligned in their weak interactions according to their masses, i.e.,  the top quark is aligned with the bottom quark and similarly for other generations?  Note that the flavor is defined by the $W^\pm$ interaction prior to symmetry breaking and the masses and mixings of fermions are determined by the Yukawa couplings. It is not apriori guaranteed that the quarks that appear in the gauge interaction will remain dominantly coupled to each other after symmetry breaking. We will call this the ``flavor alignment" puzzle. Since the Yukawa coupling matrices that determine the alignment are arbitrary in the SM, with no correlation to masses, flavor alignment is not explained.  It is therefore interesting to search for models where flavor alignment arises naturally for quarks. As for the lepton sector, the question of analogous alignment is not clear yet. For example, if the neutrino mass hierarchy is normal, it will be similar to the case of quarks although deviations from alignment are substantial due to the large mixing angles. 

There have been several proposals to address this issue: within the framework of  Randall-Sundrum models, even if the quarks of different flavors are misaligned in the bulk~\cite{kaustubh}, warping can lead to alignment at the branes; other examples include implementing an extra local $B-L$ symmetry exclusively for the third generation~\cite{Alex}, and using appropriate global symmetries~\cite{branco} that align $t$ and $b$ naturally.  In this note, we propose an alternative solution based on a combination of $U(1)_{PQ}$ symmetry~\cite{PQ} and the left-right symmetry ~\cite{LR} which resolves this puzzle for all the three generations, while simultaneously solving the strong CP, neutrino mass and dark matter problems. Flavor alignment emerges in this model in a  natural manner. The quark mixings vanish at the tree level and arise at the one loop level~\cite{bala, others} providing an explanation for their smallness. Left-right symmetry is essential for our framework since it puts the right-handed up and down quarks together. The neutrino mixings arise at the tree level via the seesaw mechanism which explains why their mixings are ``large". We emphasize that the $U(1)_{PQ}$ symmetry is needed in addition to left-right symmetry to guarantee alignment. The PQ symmetry, of course, solves the strong CP problem and provides the axion as a dark matter candidate. 

Getting quark mixings out of one loop effects requires additional colored scalars in the model~\cite{bala} with masses in the 10 TeV range. The colored scalars could either be triplets or sextets. We pursue the triplet alternative here, which is more minimal, although our discussion also applies to the case with sextets. The assignment of PQ charges guarantees that the proton decay is forbidden despite low mass for the color triplets. Since the color triplet scalar connects the up quark to the down quark, the flavor changing neutral current constraints on their couplings are much weaker.  We also give a fit for the lepton sector and find that fitting the neutrino oscillation observations and the charged lepton mass spectra requires the $W_R$ mass to be in the 20 TeV range.

The paper is organized as follows: in sec. 2, we discuss the details of the model; in sec. 3, we discuss symmetry breaking; in sec.4 we show how the flavor alignment emerges naturally as a result of the symmetries of the model. Sec. 5 is devoted to showing how the quark mixings arise from radiative effects and how they can fit the observations; sec. 6 is devoted to a demonstration of how lepton mixings arise. In sec. 7, we discuss some phenomenological implications of the model and give our conclusions. In the appendix, we explain in detail how the various stages of the symmetry breaking emerge.

\section{2. The model} 
The left-right symmetric model is based on the gauge group $SU(2)_L\times SU(2)_R\times U(1)_{B-L}$ with fermion assignments as doublets of the left and right $SU(2)$'s as
$Q^T_{L,a} = \left(\begin{array}{cc}u_L& d_L \end{array}\right)_a ;
Q^T_{R,a} = \left(\begin{array}{cc}u_R&d_R \end{array}\right)_a  ;
  \psi^T_{L,a}  =  \left(\begin{array}{cc}\nu_L &e_L \end{array}\right)_a ; 
\psi^T_{R,a} =  \left(\begin{array}{cc} N_R &  e_R \end{array}\right)_a$ 
where $a=1,2,3$ represents the family index. We choose the following Higgs multiplets: bi-doublet $\phi(2,2,0)$; $\Delta_R(1,3,+2)$ and $\Delta_L(3,1,+2)$, as in the minimal LR model ~\cite{MS}. In addition, we add the color triplet fields $\omega_{L,R}(1,1,-\frac{2}{3})$, and gauge neutral fields $\sigma_{1,2}(1,1,0)$. The former radiatively induces small CKM mixings while the latter spontaneously break the PQ symmetry. The PQ charges and gauge assignments for all the fields are presented in the Table \ref{charge}.

The most general gauge-invariant Yukawa couplings allowed by the PQ charges are given as
\begin{eqnarray}\label{yuk}
{\cal L}_Y~&=&~h_q\bar{Q}_L\phi Q_R+h_\ell\bar{\psi}_L\tilde\phi\psi_R\\\nonumber
&+& g_{ab} [Q^T_{L,a}\tau_2\omega_L C^{-1}Q_{L, b}+L\to R]\\\nonumber
&+&f_{ab}[\psi_{L, a}^T\tau_2C^{-1}\Delta_L\psi_{L, b}+ L\to R]+h.c.
\end{eqnarray}
The PQ charges by nature are asymmetric between left and right chiral fermion fields.
Note that the conjugate bi-doublet field $\tilde{\phi}\equiv \tau_2\phi^*\tau_2$ has PQ charge $-2$ and is therefore forbidden from coupling to the quarks instead it (and not $\phi$) couples to leptons. As a result, the quark Yukawa matrix $h_q$ can be diagonalized by a change of basis. This feature of our model plays a crucial role in determining the flavor alignment for quarks and is different from the minimal LRSM where both $\phi$ and $\tilde{\phi}$ couple to fermions, thereby spoiling alignment. 
The $U(1)_{PQ}$ also provides the additional advantage of solving the strong CP problem and yielding a dark matter candidate of the universe in the form of axion. Note that the coupling matrices $g_{ab}$ are general anarchic complex matrices which at the one-loop level provide quark mixings as well as the CKM CP phase.  Similarly, $f_{ab}$ being anarchic provides the lepton mixings at the tree level via the seesaw mechanism. We wish to point out that while we have used $U(1)_{PQ}$ symmetry, any global symmetry that prevents the $\tilde\phi$ coupling to quarks would also lead to the same result.
\begin{table}
\begin{tabular}{c c c c c c}
\hline
\hline
Fields & $SU(3)_C$ & $SU(2)_L$&$SU(2)_R $& B-L & PQ\\
\hline
 \\
 $ Q_{L,a}$ &   $\underline{3} $ &$ \underline{2} $ & $ \underline{1} $  &$\frac{1}{3}$    & +1  \\
 $ Q_{R,a}$ &  $ \underline{3} $ &  $\underline{1} $ & $\underline{2} $  & $\frac{1}{3}$   & -1  \\
 $\psi_{L,a}$ &  $\underline{1} $  & $\underline{2}$  & $ \underline{1}$   & $-1$   & -2  \\
 $\psi_{R,a}$&   $\underline{1}$  &  $\underline{1}$  &  $\underline{2}$  &  $-1$  & 0  \\ 
 \hline
 \\
 $\phi$& \underline{1}& $\underline{2}$ & $\underline{2}$ & $0$ & $+2$\\
  $\tilde\phi$& \underline{1}& $\underline{2}$ & $\underline{2}$ & $0$ & $-2$\\
 $\Delta_L$& \underline{1} & $\underline{3}$ & $\underline{1}$ & $+2$ & $+4$\\
 $\Delta_R$&\underline{1} & $\underline{1}$ & $\underline{3}$ & $+2$ & $0$\\
 $\omega_L$& \underline{3}& $\underline{1}$ & $\underline{1}$ & $-\frac{2}{3}$ & $-2$\\
 $\omega_R$&\underline{3} & $\underline{1}$ & $\underline{1}$ & $-\frac{2}{3}$ & $+2$\\
 $\sigma_1$&\underline{1} & $\underline{1}$ & $\underline{1}$ & $0$ & $+1$\\
 $\sigma_2$&\underline{1} & $\underline{1}$ & $\underline{1}$ & 0 & $+2$\\
\hline
\hline
\end{tabular}
\caption{The assignments of fermion and scalar fields in the model among various irreps of gauge group.}
\label{charge}
\end{table}



\section{ 3. Symmetry breaking}

To discuss the symmetry breaking pattern, we begin with the general gauge and L-R symmetric renormalizable Higgs potential:
\begin{eqnarray}
\label{HP}
 V(\phi,\Delta_{L,R}, \sigma_{1,2}, \omega_{L,R})=V_0(\phi)+V_0(\Delta_L)+V_0(\Delta_R)+V_0(\sigma_2)+V_0(\sigma_1)+V_0(\omega_L)+V_0(\omega_R)\nonumber\\
+\sum_{\alpha,\beta}V_{\alpha\beta} 
+\mu_{12} \sigma_2\sigma^{*2}_1
+\alpha_{\phi\sigma} {\rm Tr}(\phi^\dagger\tilde{\phi})\sigma^2_2+\alpha_3 Tr(\phi^\dagger\phi\Delta_R\Delta^\dagger_R)+ \beta Tr (\phi^\dagger \Delta_L\tilde\phi \Delta^\dagger_R)
\nonumber\\+\alpha_{\omega \sigma}\omega^\dagger_L\omega_R \sigma_2^{*2}
~+~L\leftrightarrow R.
\end{eqnarray}
where  $V_0$ typically contains a bilinear and quartic terms in the corresponding field, i.e., for a generic field $H$, we have
\begin{eqnarray}\label{vo}
 V_0(H)=\pm \mu^2_H H^\dagger H+\lambda_H (H^\dagger H)^2.
 \end{eqnarray}
 We choose the $+$ term in eq.(\ref{vo}) for $\omega_{L,R}$ and $\sigma_2$ and $-$ for the rest of the scalar fields. Note that some fields like $\phi$ can have more than one quartic term.
 Similarly, $V_{\alpha \beta}$  contain mixed quartic couplings between different scalar fields of the form 
 \begin{eqnarray}
 V_{\alpha\beta}=\lambda_{\alpha\beta} (H^\dagger_\alpha H_\alpha)(H^\dagger_\beta H_\beta)
 \end{eqnarray}
 The interlocked terms $\alpha_3, \beta, \alpha_{\phi\sigma}~ \text{and} ~\alpha_{\omega \sigma}$ are shown explicitly since each of them have important implications for our discussion. We also define L-R symmetry for $\sigma$ fields as $\sigma_{1,2} \to \sigma*_{1,2} $ while using the usual definitions of parity transformation for other fields, i.e., $\phi\to \phi^\dagger$ 
  for the bi-doublet and $\Delta_L\to \Delta_R$ for the triplets. All the parameters of the potential except $\alpha_3$ are real.

The minimum of the above Higgs potential leads to following Higgs field vacuum expectation values (VEV) which spontaneously break the original gauge symmetry and the PQ symmetry.
\begin{equation}\label{VEV}
\langle{\phi}\rangle=\left(
\begin{array}{cc}
 \kappa  & 0 \\
 0 & \kappa' \\
\end{array}
\right),~\langle{\Delta_{L,R}}\rangle=\left(
\begin{array}{cc}
0  & 0 \\
 v_{L,R} & 0 \\
\end{array}
\right),~ 
\begin{array}{c}
\langle{\sigma_1}\rangle=v_{PQ}, \\
\langle{\sigma_2}\rangle=v_{2}.~~ \\
\end{array}\end{equation}

This happens in several steps with the highest scale being the PQ scale $v_{PQ}$ which we choose to be $\sim 10^{12}$ GeV for a phenomenologically viable axion dark matter~\cite{jekim}. 
The next scale in the model is the right-handed scale, $v_R $, which is chosen to be of the order of 50 TeV to be compatible with the rare process constraints. The SM electroweak symmetry breaking takes place when the fields $\phi$ acquire a VEV such that $\kappa ^2 + \kappa'^2 = v^2_{wk}$.  
Getting the electroweak VEV to be of the order of a 100 GeV requires a fine tuning between the $\mu^2_\phi$ and  $\lambda_{\phi \sigma_1}v^2_{PQ}$ which is the usual fine tuning of the invisible axion models~\cite{jekim}. The value of the coupling $\beta$ is taken to be $\sim 10^{-7}$ to protect the neutrino masses from receiving large type-II seesaw contributions. In the appendix, we discuss in detail how this symmetry breaking pattern arises.

 
 From the discussion above, we find that the axion field has components in all the Higgs fields except the $\Delta_R^0$. This is expected since $\Delta_R$ is PQ neutral. The contribution from $\Delta_L^0$ is small since if $v_L$ is small compared to other VEVs.  For leading order in $\kappa'\ll \kappa$, we get
\begin{eqnarray}
{\it a}={\cal N}\left(\frac{4\kappa^{'2}}{\kappa v_{PQ}}\chi_{\phi_1}+\frac{4\kappa'}{v_{PQ}}\chi_{\phi_2}-\frac{2v_2}{v_{PQ}}\chi_{\sigma_2}+ \chi_{\sigma_1}\right)
\end{eqnarray}
 where $\chi_H$ is used to denote the imaginary part of the corresponding complex field $H$ and ${\cal N}$ is the normalization factor. 
 


\section{4. Flavor alignment} To see how this model naturally leads to the alignment of quark flavors, we note that without loss of generality, we can choose the basis for the quark fields before symmetry breaking such that the quark Yukawa couplings $h_q$ is diagonal. This implies that the up quarks of different generations are aligned with the down quarks of the corresponding generation. The mass of the top and bottom quarks are given at the tree level by $m_t/m_b=\kappa/\kappa' =m_c/m_s$. The second tree level relation is not very well satisfied but is fixed by the one loop correction. Since the one loop corrections are expected to be small for the second and the third generation, flavor alignment remains.  

The situation in the lepton sector is slightly different. It follows from the Yukawa couplings that Yukawa matrix $h_\ell$ can also be diagonalized by the choice of an appropriate basis. In this basis, both the charged lepton as well as the Dirac mass matrix for the neutrinos are diagonal. However, the neutrino masses arise out of seesaw mechanism which involves the right-handed neutrino mass matrix, $M_N$, which is a general $3\times3$ complex symmetric matrix unconstrained by the PQ symmetry. 

By adjusting the elements of $M_N$, we can get any flavor of neutrino to go with any flavor of charged lepton. In other words, this can allow for large departures from alignment which is the case for leptons. In fact, this can also allow for the extreme case of misalignment which occurs when there is an inverted hierarchy of neutrino masses.


\section{5. One loop corrections and fits to quark masses and mixings} 
\begin{figure}
\centering
 \includegraphics[width=0.32\textwidth]{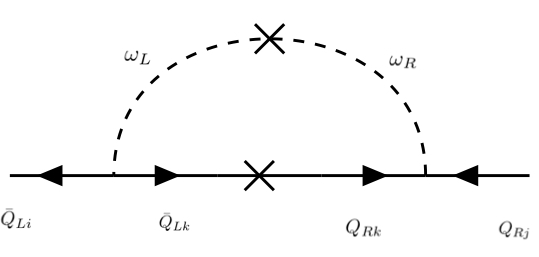}
\caption{Quark mixings arising form the one-loop contributions from color scalars $\omega_{L,R}$.}
\label{one-loop}
\end{figure}

From the Yukawa couplings in eq.(\ref{yuk}), it can be seen that the up and down quark mass matrices are proportional to each other at the tree level implying no mixing among the quark flavors. After the symmetry breaking, there is a new tree level contribution to the quark mass matrices coming from the $\alpha_{\phi\sigma}$ term generating an effective coupling of the form $h_q \bar{Q}_L\tilde\phi Q_R$ but the generation structure of the coupling is the same as the coupling $h_q$ and can, therefore, be absorbed in it by a redefinition. However, at the one-loop level the color triplets, $\omega_{(L,R)}$, generate the necessary quark mixing through radiative corrections as shown in fig. \ref{one-loop}. 
To study this, we do an orthogonal rotation of $\omega_{(L,R)}$ by an angle $\alpha$ to go the mass basis $\omega_{(1,2)}$ with masses $m_{1,2}$ respectively. The mass matrices for up and down quark masses including the one loop contribution (from figure \ref{one-loop}) can be written as  (in the limit
 $m_{1,2}\gg m_{top}$)
\begin{eqnarray}
M^{u}_{ij}&=&h_q\kappa+\frac{3 \sin 2\alpha}{16\pi^2} \ln\frac{m_1}{m_2}(g^{\dagger}h_{q }g)_{ij}\kappa',\\\nonumber
 M^{d}_{ij}&=&h_q\kappa'+\frac{3 \sin 2\alpha}{16\pi^2} \ln\frac{m_1}{m_2}(g^{\dagger}h_{q }g)_{ij}\kappa.
\end{eqnarray}
where the first terms in both these expressions denote the tree level contribution and the second, the one loop one. It is clear that the up and down quark mass matrices are not proportional to each other anymore and generate nonzero CKM angles.
The phases appearing in the diagonal elements of the complex-symmetric Yukawa matrix $g$ can be absorbed through a redefinition of quark fields. In the case of color triplets, the one-loop corrections to up-type quarks are proportional to the tree-level masses of down-type quarks. This feature can naturally explain the inverted hierarchy of masses for first generation quarks since the tree level masses of the first generation quarks are much smaller than the one loop effect (which has the property of inverting them). For the charm and strange quarks, on the other hand, no such inversion takes place since the tree level contribution to charm mass is already close to observed value.

Since the color triplet masses are free parameters, we take $m_1=e\times m_2\sim10~$TeV and a mixing angle $\alpha=\pi/4$ for fitting the fermion masses. This choice is however not essential to get a good fit for the masses. 
A point in the parameter space which fits the fermion masses and mixing parameters are given below.
We find that the neutrino mass fit requires the right-handed scale $v_R \geq 50$ TeV.  Recalling that  $g_L\simeq 0.65$ and the lowest value of $g_R$ allowed in the LR model is $g_R \geq 0.55 g_L$~\cite{DMZ}, we find that for our model to work, we must have an $M_{W_R}\geq 18$ TeV.

The quark sector input parameters:
\begin{eqnarray}
h_q = diag\{6.9914\times 10^{-6},~3.3811\times 10^{-3},~0.9335 \},\kappa/\kappa'=59.24~~\kappa=173.98 ~GeV\nonumber\\ 
g=\left(
\begin{array}{ccc}
 0.6861 & 0.7081-0.0007 i & 0.0082 - 0.0260 i \\
0.7081-0.0007 i & 0.7255 & 0.1236 - 0.0007 i \\
0.0082 - 0.0260 i & 0.1236 - 0.0007 i  & 0.3076 \\
\end{array}
\right)
\end{eqnarray}
lead to 
\begin{eqnarray}
 m_u=~~1.35~MeV, ; m_d=~~4.83~MeV; m_c=0.589~GeV; \nonumber\\
m_s=~61.15~MeV;m_b=3.04~GeV; m_t=162.42~GeV
\end{eqnarray}
\begin{eqnarray}
|V_{CKM}|= \left(
\begin{array}{ccc}
 0.9743 & 0.2254 & 0.0036 \\
0.2253 & 0.9735 & 0.0406 \\
0.0088 & 0.0398 & 0.9992 \\
\end{array}
\right);~~~ \text{and}~J_{CP}=3.07\times10^{-5};
\end{eqnarray}

These numbers correspond to a point in the parameter space that can accommodate all the fermion masses and mixings.We have chosen the values of the masses in the $\overline{MS}$ scheme corresponding to the renormalization scale $\mu=m_t(m_t)$. We do not claim that this set of predictions are generic to the model but is given to show that our basic idea works in giving a realistic model.

\section{6. Lepton sector:} In the lepton sector, the Yukawa coupling $h_\ell$ can be rotated into a real diagonal matrix, like in the quark sector. The couplings $h_\ell$ is now completely determined by the charged lepton masses such that $h_\ell = \{m_e, m_\mu, m_\tau \}/\kappa$. The neutrino masses are generated through a combination of type-I and type-II seesaw mechanisms and the neutrino mixings arise from the right-handed neutrino mass matrix which is completely anarchic. The light neutrino mass matrix can be approximated as
$M_\nu=f v_L - \frac{h_\ell f^{-1} h_\ell}{v_R} \kappa'^2$.
The diagonal phases of the complex-symmetric Yukawa coupling $f$ can be absorbed through a re-definition of neutrino fields. 

Lepton sector input parameters:
\begin{eqnarray}
\frac{f}{10^{-2}}
=\left(
\begin{array}{ccc}
 0.11438  & 1.7757+ ~0.030602 i & -7.28103 + 0.0871 i \\
1.7757+ ~0.030602 i & ~14.6833& -99.81152 + ~1.82564 i\\
-7.28103 + 0.0871 i&  -99.81152 + ~1.82564 i& ~99.73409 \\
\end{array}
\right)\nonumber\\
v_L =0.5501~eV,~v_R = 50~TeV
\end{eqnarray}
Output in the lepton sector:
\begin{eqnarray}
\sin^2\theta_{13}=0.0215; 
\sin^2\theta_{12}= 0.3209;
\sin^2\theta_{23}=0.43~~\nonumber~\\
m_1 = 0.00294~eV; m_2=0.00917~eV; m_3=0.05058~eV;
\sin\delta_{CP}=-0.951
\end{eqnarray}

This leads to
\begin{eqnarray}
 \Delta m^2_{21}=7.56\times10^{-5}~eV^2;
 \Delta m^2_{31}=2.55\times10^{-3}~eV^2;
\nonumber\\
 \sum m_\nu =~~~0.062701 ~eV~and~ \langle m_{\beta\beta}\rangle=0.36\times10^{-3}eV
\end{eqnarray}

\section{7. Discussions and conclusion:}
Several points worth noting about the model:

\subsection{ 7a. Flavor changing neutral current constraints} There are two sources of quark flavor violation in the model: 

(i) As in the minimal LR model, there are two neutral Higgs-mediated flavor changing effects, once the one loop effects are included to generate quark mixings. This puts a lower bound on the $\phi^0_2$ mass in the range of 10 TeV or higher. This is lower than the $v_R$ scale in our model and is natural.

(ii) A second source of flavor violation in the model comes from the quark couplings to the color triplet fields $\omega_{L,R}$. The corresponding case for color sextet fields were analyzed in Ref.~\cite{fortes}. The difference in our case is that color triplets always connect up quarks to down quarks, which is similar to the the up-down connecting sextet field analyzed in ~\cite{fortes}. At the tree level, they lead to flavor violating decays 
such as $B\to \pi\pi$, $K\to \pi\pi$ etc. From Ref.~\cite{fortes}, we find that in the case products such as $g_{12}g^*_{11}$ etc are bounded and for $M_{\omega}\sim 1$ TeV, these bounds are of order one. Since in our case we take $m_{1,2}\sim 10$ TeV, these bounds are weaker and are consistent with our choice of $g_{ab}$. The second type of FCNC comes from box graphs which lead to processes such as $K-\bar{K}$, $B-\bar{B}$ mixing. Again, the most stringent constraint of this type are:
$\sum_a g_{1a}g^*_{a2}\leq 0.1(M_{\omega}/TeV) $ from $K-\bar{K}$ mixing and $\sum_a g_{1a}g^*_{a3}\leq 0.2(M_{\omega}/TeV) $ from $B_d-\bar{B}_d$ mixing and $\sum_ig_{2a}g^*_{a3}\leq  1.0(M_{\omega}/TeV) $ from $B_s-\bar{B}_s$ mixing. Clearly for our choice of $M_{\omega}\sim 10$ TeV, our choice of parameters are quite consistent with these bounds.
Since the phases in our model are small and the triplet mass is 10 TeV, we find that the correction to the CP violating effects in the meson systems is consistent with observations. According to our fermion mass fit, we expect that corrections to standard model predictions for FCNC effects should appear soon. The details regarding this are presently under study.

(iii) As far as the leptonic flavor change is concerned, the dominant contributions come from $\Delta_{L,R}$ exchange at the one-loop level for $\mu\to e+\gamma$ etc. and at the tree level for $\mu\to 3 e$ and $\tau\to 3\ell$ decays. For $\Delta$ masses near 50 TeV, our parameter choice is consistent with current bounds from these as well as other processes.

(iv) Any symmetry that forbids the $\tilde{\phi}$ coupling to quarks will do for achieving alignment. We choose PQ symmetry since it not only helps us in achieving that goal but also provides a solution to the strong CP problem.

\subsection{7b. Further comments}

\begin{itemize}
\item The choice of our PQ charges for the lepton sector is dictated by the requirement that proton decay be forbidden since all scales except the PQ scale are in the multi-TeV range.



\item The lightest right-handed neutrino has a mass around 100 GeV and is coupled to all three charged leptons. However, its production rate is suppressed due to the heavy $W_R$ mass as well as due to a small heavy-light neutrino mixing, with the largest mixing being the $V_{eN_e}\sim 10^{-6}$. It is therefore not observable at the LHC. 


\item We have not explored the question of leptogenesis in the model; however, we note that there are two right-handed neutrinos which are quasi-degenerate in the model, which is a pre-requisite for leptogenesis and also that $M_{W_R}>18 $ TeV, which guarantees that all washout effects are small~\cite{hambye}.

\item Our fit gives a  $\sum m_\nu = 0.06$ eV. This is below the current bound from Planck and other experiments ~\cite{review}. However, this can be tested in forthcoming experiments such as the LSST survey and EUCLID mission etc. which are expected to bring it down to 0.02 eV.

\item Our model can accommodate the current central value for the leptonic CP phase~\cite{Nova}. 

\end{itemize}

In conclusion, we have presented a simple resolution of the ``flavor alignment puzzle" of the standard model using a combination of left-right with Peccei-Quinn symmetry. The model also solves the strong CP problem as well as the problem of neutrino masses and mixings. 


\section*{Appendix}
In this appendix, we explain how the VEVs chosen in the model arise from the Higgs potential minimization. We seek a minimum with the  VEV hierarchy $v_{PQ}\gg v_R \gg \kappa\sim\kappa'\sim v_2\gg v_L$ where these VEVs are defined in the eq.(\ref{VEV}). We will proceed from the highest scale and explain how the VEVs arise at each scale using the language of effective renormalizable potential at that scale.

The highest scale in the model is the PQ scale denoted by $v_{PQ}$ and it arises from the effective potential given by:
\begin{equation}
V(\sigma_1)=-\mu^2_{\sigma_1}\sigma^*_1\sigma_1+\lambda_{\sigma_1}(\sigma^*_1\sigma_1)^2\nonumber
\end{equation}
in the usual way. With a choice of $\mu_{\sigma_1}\sim 10^{12}$ GeV, we get the desired $v_{PQ}$ for the invisible axion.

Now, we will consider the potential for the other PQ charged Higgs field $\sigma_2$ given by
\begin{equation}
V(\sigma_2)=\mu^2_{\sigma_2}\sigma^*_2\sigma_2+\lambda_{\sigma_2}(\sigma^*_2\sigma_2)^2 +\mu_{12} \sigma_2\sigma^{*2}_1\nonumber
\end{equation}
where $\mu_{\sigma2}$ is naturally chosen to be $\sim v_{PQ}$. For this potential, the spontaneous symmetry breaking is solely due to the cubic term in the potential. For our desired hierarchy, 
the minimization gives $v_2=\frac{\mu_{12}v^2_{PQ}}{\mu^2_{\sigma_2}}\sim \mu_{12}$. It should be noted that a small value of $\mu_{12}$ is radiatively stable since it is the only term in the Lagrangian that softly breaks the $Z_{2 \sigma_2}$ ( $\sigma_2\leftrightarrow -\sigma_2$). This allows us to naturally choose a value for $v_2 \sim\mu_{12}\sim v_{wk}$.

The next scale in the model is the parity breaking scale $v_R$ which arises similarly to the $v_{PQ}$ from the effective potential for $\Delta_R$. In this case, the effective mass term for the $\Delta_R$ field after integrating out the $\sigma_1$ field has the symbolic form:
\begin{equation}(-\mu^2_{\Delta_R}+\lambda_{\sigma_1\Delta_R}v^2_{PQ})Tr(\Delta^\dagger_R\Delta_R)+\lambda_{\Delta_R}Tr(\Delta^\dagger_R\Delta_R)^2.\nonumber
\end{equation}
Since $v_R\ll v_{PQ}$, we need a fine tuning between the two contributions ($-\mu^2_{\Delta_R}~\text{and} ~\lambda_{\sigma_1\Delta_R}v^2_{PQ} $) to the $\Delta_R$ mass term. This is one of the fine tunings we referred to in the text.

Finally, we address the SM breaking VEVs $(\kappa$ and $\kappa'$). Note that at this scale $SU(2)_R\times U(1)_{B-L}$ is already broken and the fields acquiring VEVs are given by $\phi^0_1$ and $\phi^0_2$, which are the two  neutral members of the complex bi-doublet $\phi$ in our model. Since $\kappa,\kappa'\ll v_R$, we write the effective potential for the $\phi^0_{1,2}$ using breaking scales $v_{PQ}$ and $v_R$. We will first set $\alpha_{\phi\sigma}=0$. In this case, the potential for the fields $\phi^0_{1,2}$ becomes,
\begin{eqnarray}
V(\phi^0_1, \phi^0_2)=(-\mu^2_\phi+\lambda_{\Delta_R\phi}v^2_R+\lambda_{\sigma_1\phi}v^2_{PQ})(\phi^{0*}_1\phi^0_1+\phi^{0*}_2\phi^0_2)+\lambda_{eff}(\phi^{0*}_1\phi^{0}_1+\phi^{0*}_2\phi^{0}_2)^2\nonumber\\+\alpha_3 v^2_R(\phi^{0*}_2\phi^{0}_2)
.\nonumber
\end{eqnarray}
We need some fine tuning between $\lambda_{\Delta_R\phi}v^2_R$ and $\lambda_{\sigma_1\phi}v^2_{PQ}$ to get a weak scale mass for $\phi$. A convenient way to find the minimum of this potential is by parameterizing $\langle\phi^0_1\rangle=\kappa=r \cos\theta$ and $\langle\phi^0_2\rangle=\kappa'=r \sin\theta$ and rewriting the potential as
\begin{eqnarray} 
V(r, \theta)= -\mu^2_{eff}r^2+\lambda_{eff}r^4+\alpha_3 v^2_R r^2{\rm sin}^2\theta.\nonumber
\end{eqnarray}
Clearly, the extremum of this potential corresponds to $\sin\theta=0$ or $\kappa'=0$ for $\alpha_3> 0$ (which is what we choose).

The question now is how can a non-zero $\kappa'$ be induced? For this we look towards the contribution of $\sigma_2$ VEV to the $\phi^0_{1,2}$ potential. Once the VEV $<\sigma_2>\neq 0$ it induces a mixing between $\phi^0_1$ and $\phi^0_2$ from the   term  $\alpha_{\phi\sigma} {\rm Tr}(\phi^\dagger\tilde{\phi})\sigma_2\sigma_2$. The effective potential in the presence of this term becomes
\begin{eqnarray}
V_{eff}(\phi^0_1, \phi^0_2)=(-\mu^2_\phi+\lambda_{\Delta_R\phi}v^2_R+
\lambda_{\sigma_1\phi}v^2_{PQ})(\phi^{0*}_1\phi^0_1+\phi^{0*}_2\phi^0_2)+\lambda_{eff}\left(\phi^{0*}_1\phi^{0}_1+\phi^{0*}_2\phi^{0}_2\right)^2\nonumber\\
+\alpha_3 v^2_R(\phi^{0*}_2\phi^{0}_2)+\alpha_{\phi\sigma} v^2_2\phi^0_1\phi^0_2+h.c. \nonumber
\end{eqnarray}

In terms of $r$ and $\theta$, we get 
$$V(r, \theta)= -\mu^2_{eff}r^2+\lambda_{eff}r^4+\alpha_3 v^2_R r^2{\rm sin}^2\theta+\alpha_{\phi\sigma} v^2_2r^2\sin\theta\cos\theta$$
Minimizing this with respect to $\theta$, we get 
$$ \alpha_3 v^2_R r^2\sin 2\theta+\alpha_{\phi\sigma} v^2_2 r^2 \cos 2\theta =0$$ which leads to a non-zero minimum for $\theta$ at $\tan 2\theta=\frac{\alpha_{\phi\sigma} v^2_2}{\alpha_3 v^2_R}$. By choosing appropriate signs for $\alpha_{\phi\sigma} $ and $\alpha_3 $, we can obtain $\theta<\pi/4$ which gives the phenomenologically preferred hierarchy: $\kappa>\kappa'\sim v_{wk}$ . 

The potential for $v_L$ is given approximately as
$$\lambda_{\sigma_1\Delta_L}v^2_{PQ}Tr(\Delta_L^{\dagger}\Delta_L) + \kappa'^2 v_R\Delta_L^0.$$ Minimizing this, we get $v_L\sim \kappa'^2 v_R/(\lambda_{\sigma_1\Delta_L}v^2_{PQ})$ which is  very small compared to other VEVs as desired.

\section*{ Acknowledgements} We thank K. Agashe and K. S. Babu or discussions and comments on the manuscript. This work is supported by the US National Science Foundation under Grant No. PHY1620074.


\begin{thebibliography}{99}


\bibitem{kaustubh} See for example, T.~Gherghetta and A.~Pomarol,
  Nucl.\ Phys.\ B {\bf 586}, 141 (2000); K.~Agashe, G.~Perez and A.~Soni,
  Phys.\ Rev.\ D {\bf 71}, 016002 (2005); 

 

\bibitem{Alex}   K.~S.~Babu, A.~Friedland, P.~A.~N.~Machado and I.~Mocioiu,
  JHEP {\bf 1712}, 096 (2017).
  
  \bibitem{branco}  F.~J.~Botella, G.~C.~Branco, M.~Nebot, M.~N.~Rebelo and J.~I.~Silva-Marcos,
  Eur.\ Phys.\ J.\ C {\bf 77}, no. 6, 408 (2017).
   

\bibitem{PQ} R.~D.~Peccei and H.~R.~Quinn,
  Phys.\ Rev.\ Lett.\  {\bf 38}, 1440 (1977).
  
   \bibitem{LR} J.C. Pati and A. Salam, Phys. Rev. D {\bf 10}, 275 (1974); 
R. N. Mohapatra and J. C. Pati, Phys. Rev. D {\bf 11}, 566, (1975);  R. N. Mohapatra and J. C. Pati, Phys. Rev. D {\bf 11}, 2558 (1975); 
G. Senjanovi\'{c} and R. N. Mohapatra, Phys. Rev. D {\bf 12} 1502 (1975). 

  
  \bibitem{MS} R. N. Mohapatra and G. Senjanovic, Phys. Rev. Lett {\bf 44}, 912 (1980).
  
  \bibitem{jekim} J. E. Kim, Phys. Rev. Lett. 43, 103 (1979); M. A. Shifman, A. I. Vainshtein, and V. I. Zakharov, Nucl. Phys. B 166, 4933 (1980) [[22] A.P. Zhitnitsky, Yad.Fiz. (1980) 31: 497 Sov. J. Nucl. Phys. (1980) 31: 260];  M. Dine, W. Fischler and M. Srednicki, Phys. Lett. B
104, 199 (1981). For reviews, see J. E. Kim,  Phys.Rep. 150, 1 (1987); J.~E.~Kim, S.~Nam and Y.~K.~Semertzidis,
  Int.\ J.\ Mod.\ Phys.\ A {\bf 33}, no. 03, 1830002 (2018).
  
  \bibitem{bala} B.~S.~Balakrishna,
  Phys.\ Rev.\ Lett.\  {\bf 60}, 1602 (1988); B.~S.~Balakrishna, A.~L.~Kagan and R.~N.~Mohapatra,
  Phys.\ Lett.\ B {\bf 205}, 345 (1988); K.~S.~Babu, B.~S.~Balakrishna and R.~N.~Mohapatra,
  Phys.\ Lett.\ B {\bf 237}, 221 (1990).
  
      
 \bibitem{others} Here is an incomplete list of references to litrerature: K.~S.~Babu and E.~Ma,
  Mod.\ Phys.\ Lett.\ A {\bf 4}, 1975 (1989);
 X.~G.~He, R.~R.~Volkas and D.~D.~Wu,
  Phys.\ Rev.\ D {\bf 41}, 1630 (1990);
  Z.~G.~Berezhiani and R.~Rattazzi,
  Phys.\ Lett.\ B {\bf 279}, 124 (1992);
 B.~A.~Dobrescu and P.~J.~Fox,
  JHEP {\bf 0808}, 100 (2008); P.~W.~Graham and S.~Rajendran,
  Phys.\ Rev.\ D {\bf 81}, 033002 (2010);
A.~Ibarra and A.~Solaguren-Beascoa,
  Phys.\ Lett.\ B {\bf 736}, 16 (2014);
 M.~Baumgart, D.~Stolarski and T.~Zorawski,
  Phys.\ Rev.\ D {\bf 90}, no. 5, 055001 (2014);  For a review, see K.~S.~Babu,
  arXiv:0910.2948 [hep-ph];
A.~E.~C�rcamo Hern�ndez, S.~Kovalenko and I.~Schmidt,
  JHEP {\bf 1702}, 125 (2017).
 
 
  \bibitem{fortes} E.~C.~F.~S.~Fortes, K.~S.~Babu and R.~N.~Mohapatra,
  arXiv:1311.4101 [hep-ph]; K.~S.~Babu, P.~S.~Bhupal Dev, E.~C.~F.~S.~Fortes and R.~N.~Mohapatra,
  Phys.\ Rev.\ D {\bf 87}, no. 11, 115019 (2013).
  
  \bibitem{DMZ} J.~Brehmer, J.~Hewett, J.~Kopp, T.~Rizzo and J.~Tattersall,
  JHEP {\bf 1510}, 182 (2015); P.~S.~B.~Dev, R.~N.~Mohapatra and Y.~Zhang,
  JHEP {\bf 1605}, 174 (2016).
  
 \bibitem{Nova} P.~Adamson {\it et al.} [NOvA Collaboration],
  Phys.\ Rev.\ Lett.\  {\bf 118}, no. 23, 231801 (2017)
  [arXiv:1703.03328 [hep-ex]]; K. Abe et al. [T2K Collaboration], Phys. Rev. {\bf D 91}, no. 7, 072010 (2015) [arXiv:1502.01550 [hep-ex]].
  
  
  \bibitem{hambye} J.~M.~Frere, T.~Hambye and G.~Vertongen,
  JHEP {\bf 0901}, 051 (2009);
  P.~S.~Bhupal Dev, C.~H.~Lee and R.~N.~Mohapatra,
  J.\ Phys.\ Conf.\ Ser.\  {\bf 631}, no. 1, 012007 (2015).
  
  \bibitem{review} P. A. R. Ade et al. [Planck Collaboration], Astron. Astro- phys. 594, A13 (2016) [arXiv:1502.01589 [astro-ph.CO]]; for a review, see S. Hannestad and T. Schwetz, JCAP 1611, no. 11, 035 (2016) [arXiv:1606.04691 [astro-ph.CO]].
  
\end{thebibliography}
\end{document}